# First-principles studies on electrical resistivity of iron under pressure

Xianwei Sha[*] and R. E. Cohen

Carnegie Institution of Washington, 5251 Broad Branch Road, NW, Washington, D. C. 20015, U. S. A.

**Abstract**

We investigate the temperature and pressure dependences of the electrical resistivity for bcc and hcp Fe using the low-order variational approximation and theoretical transport spectral functions calculated from first principles linear response linear-muffin-tin-orbital method in the generalized-gradient approximation. The calculated values are in close agreement with available experimental data, and show strong increase with temperature and decrease with pressure. We also discuss the behavior of the electrical resistivity for the bcc → hcp phase transition.

PACS number(s): 72.15.Eb, 72.10.Di, 63.20.kd, 63.70.+h, 71.15.Mb

[*] Now at the National Institute of Standards and Technology, 100 Bureau Drive, Gaithersburg, MD 20899.

## I. Introduction

Iron is one of the most abundant elements on the Earth, and is essential in many technological applications. Its electronic transport properties at high pressures and temperatures control the signature of magnetic processes in the core of the Earth, and have been used to understand the phase transitions in iron. For example, the transition in iron from the ambient ground-state bcc structure to the high-pressure hcp phase was identified via mapping out the pressure dependence of electrical resistance during shock compression.[1-3] Garg et al. reported an abrupt increase of the electrical resistance during the phase transition in measurement of iron in a diamond anvil cell up to 25 GPa in quasihydrostatic conditions.[4] Recently new experimental techniques have been developed to measure the electrical resistivity at much higher pressure and temperature regimes. Bi et al. reported static measurements of electrical resistivity of iron under shock compression up to 208 GPa using iron sample encapsulated in a single-crystal sapphire cell.[5,6] Stacey and Loper estimated the electrical and thermal conductivities of iron at high pressure and the Earth's core conditions.[7] There are still some fundamental problems, such as understanding the pressure dependence of the electrical resistivity and the origin of the sharp divergence between the static and dynamic resistivity data in iron under pressure.[3] So far, transport properties have drawn relatively little theoretical attention. Zhang and Drabold recently developed expressions for the electrical conductivity within a single-electron picture and perturbation theory from the principle of virtual work and coarse graining.[8] Savrasov and Savrasov calculated the transport properties for a number of elemental metals using the low-order variation approximation (LOVA) and theoretical transport spectral distribution functions of electron-phonon interaction obtained from *ab initio* linear response calculations, and their calculated values generally agree well with experiment.[9] As far as we know, no previous first-principles based calculations have been reported to examine the electrical resistivity in iron. Here we apply to bcc and hcp iron similar theoretical techniques as Savrasov et al., except that we use the generalized gradient approximation (GGA) for the exchange correlation functional instead of the local density approximation (LDA). This is essential since previous theoretical calculations reported that GGA shows significant improvements over LDA to accurately describe many ground-state properties for iron.[10-12] We further investigate the temperature and pressure dependences of the transport properties. We briefly introduce the theoretical and computational methods in section II, and present the results and related discussion in section III. We conclude with a summary in Section IV.

## II. Theoretical methods

We calculate the electrical resistivity $\rho$ and thermal conductivity $w^{-1}$ using the low-order variational approximation and obtain the transport spectral functions from first-principles linear response calculaitons[9, 13, 14]

$$\rho = \frac{\pi \Omega_{cell} k_B T}{N(\varepsilon_F)\langle v_x^2 \rangle} \lambda_{tr} \qquad (1)$$

$$w^{-1} = \frac{\pi k_B N(\varepsilon_F)\langle v_x^2 \rangle}{6\Omega_{cell} \lambda_{tr}} \qquad (2)$$

where $N(\varepsilon_F)$ is the electronic density of states per atom and per spin at the Fermi level, $\langle v_x^2 \rangle$ is the average square of the x component of the Fermi velocity, $k_B$ is the Boltzmann constant, T is temperature in Kelvin, and $\Omega_{cell}$ is the unit cell volume. The transport constant $\lambda_{tr}$ is defined by

$$\lambda_{tr} = 2\int_0^\infty \alpha_{tr}^2 F(\omega) \frac{d\omega}{\omega} tr \qquad (3)$$

We calculate the theoretical transport spectral function $\alpha_{tr}^2 F(\omega)$ based on the phonon dispersion and electron-phonon scattering obtained from first-principles linear response calculations. More details about the theoretical methods can be found in Ref. 9. The LOVA approximations works best at temperatures between $\Theta_{tr}/5$ and $2\Theta_{tr}$, where $\Theta_{tr}$ is the average transport frequency.[9] $\Theta_{tr}$ strongly depends on the pressure and is about 311 K for bcc Fe at ambient pressure and estimated to be about 633 K for hcp Fe at 330 GPa, the pressure of Earth's core (as shown in Table I). At very low temperatures, the electron-electron scattering, impurity scattering, and size effects become important contributions to electrical resistivity, in addition to the electron-phonon scattering considered here. At very high temperatures, anharmonicity and the Fermi surface smearing have to be taken into account.

The computational details to obtain the phonon dispersion and thermal euqation of state properties for both bcc[15, 16] and hcp Fe[17, 18] have been previously published. To briefly summarize, the approach is based on the density functional perturbation theory, using multi-$\kappa$ basis sets in the full-potential Linear-Muffin-Tin-Orbital (LMTO) method.[14, 19] The induced charge densities, the screened potentials and the envelope functions are represented by spherical harmonics inside the non-overlapping muffin-tin spheres surrounding each atom and by plane waves in the remaining interstitial region. We use the Perdew-Burke-Ernzerhof (PBE) GGA approximation for the exchange and correlation functional.[20] The **k**-space integration is performed using the improved tetrahedron method.[21] We use the perturbative approach to calculate the self-consistent change in the potential,[9, 13] and determine the dynamical matrix for a set of irreducible **q** points. We refer the readers for computational details in our previous publications.[15-18]

## III. Results and discussion

Table I lists the calculated values for the transport constant $\lambda_{tr}$ and the average transport freency $\Theta_{tr}$ at several selected volumes for both bcc and hcp Fe. The average transport frequency $\Theta_{tr} = \sqrt{\langle \omega^2 \rangle_{tr}}$ is close to the average phonon energy, and its value increases strongly with pressure (decrease in atomic volume according to the thermal equation of state[15, 17]), consistent with our previous lattice dynamics analysis which shows that the average phonon frequency increases with pressure for both bcc and hcp Fe.[15, 17] On the contrary, $\lambda_{tr}$ shows complex volume dependence, although its value in general becomes smaller at higher pressures. As shown in equation 3, $\lambda_{tr}$ is proportional to the integral of $\alpha_{tr}^2 F(\omega)/\omega$, so its value relies on the pressure dependences of both the transport spectral function $\alpha_{tr}^2 F(\omega)$ and the phonon frequency $\omega$. In comparison to the previously reported average transport frequencies for a number of elemental metals,[9] both bcc and hcp Fe have relatively large $\lambda_{tr}$ values, which makes direct comparisons of the calculated transport properties with experiment more relevant at the intermediate temperature regimes.

At a given atomic density, the thermal conductivity $w^{-1}(T)$ drops rapidly at up to 100 K for both bcc and hcp Fe, and becomes essentially temperature independent at higher temperatures, as shown in Fig. 1(a) and 2(a). At ambient pressure, the measured experimental data for bcc Fe[22-24] show similar temperature dependence, and agree within 10% with the calculated values at the ambient equilibrium volume (V=79.6 bohr$^3$/atom) in the intermediate temperature regimes. At very low temperature, the discrepancies between the calculations and experiment are relatively large mainly due to the neglect of the electron-electron scattering, the impurity scattering and the size effects in the present calculations. For bcc and hcp Fe, the thermal conductivity both increases with pressure.

The electrical resistivity of bcc and hcp Fe shows a strong linear increase with temperature at T > 50 K, as shown in Fig. 1(b) and 2(b). The ambient-pressure experimental data for bcc Fe[22-24] agree well with the calculated values at V=79.6 bohr$^3$/atom. The relatively large difference at T=400 K is partly due to the slight increase in the equilibrium volume due to thermal expansion. On the other hand, the calculated electrical resistivity strongly depends on the atomic volume. In general, the larger the atomic volume is, the larger is the electrical resistivity at the same temperature. At T > 0.5$\Theta_D$(T,V), the Bloch-Grüneisen formula is widely used to describe the electrical resistivity of a metal [5]

$$\rho = \frac{BT}{4A\Theta_D^2(T,V)} \qquad (4)$$

Where $\Theta_D$(T,V) is the Debye temperature, A is the atomic weight and B is a material constant. We previously obtained $\Theta_D$(T,V) for both bcc and hcp Fe via fitting the calculated Helmholtz free energies at different volumes and temperatures to a Debye model, and found that the Debye temperature is essentially temperature independent at T > 0.5$\Theta_D$(T,V),[15, 25] corresponding to the observed linear increase of $\rho$ with temperature at a given atomic density. The calculated $\Theta_D$(T,V) increases rapidly with pressure, thus the electrical resistivity drops rapidly according to the Bloch-Grüneisen formula, consistent with our calculated data. Previous theoretical[26, 27] and experimental[28] investigations also reported a rapid increase of the Debye temperature with pressure for hcp Fe. At a given temperature, the ratio between our calculated electrical resistivities at several different atomic densities is in close agreement with the ratio of $\frac{1}{\Theta_D^2(T,V)}$. Overall, our calculated temperature and pressure dependences of the electrical resistivity for bcc and hcp Fe agree with the Bloch-Grüneisen formula.

Bi et al. showed that the Bloch-Grüneisen formula is valid for hcp iron at pressures up to 208 GPa and temperatures up to 5220 K.[5, 6] Using the calculated thermal conductivities at the immediate temperature regime and the first-principles Debye temperatures at various pressures, we estimated the electrical resistivity at high pressures and temperatures using the Bloch-Grüneisen formula. The calculated resistivities at 100 GPa, 2010 K and 208 GPa, 5220 K are 56.5 and 88.4 μΩ*cm, respectively, in reasonable agreements with the values Bi et al. obtained at 68.9 and 130.7 μΩ*cm. Bi et al. calculated their high-temperature high-pressure electrical resistivity based on the analysis of the experimental Grüneisen parameter, where our previous calculations show that the Grüneisen parameter of hcp Fe shows very complex temperature and pressure dependences,[25] which is one the reasons for differences of the high-temperature high-pressure electrical resistivity between our results and Bi et al.'s values.

The evolution of the electrical resistivity of iron with pressure has been measured in both dynamic compression[1, 2] and diamond anvil cell[4] experiments. However, there is a sharp divergence between the static and dynamic resistivity data reported.[3] Earlier dynamic electrical resistance shows a slight increase of resistivity with pressure,[1, 2] while later experiments gave opposite trend.[3, 4] At ambient temperature, our calculated electrical resistivity for both bcc and hcp Fe slight decreases with pressure, as shown in Fig. 3. A similar slight decrease of the electrical resistivity with pressure has been previously reported for three hcp metals Ti, Zr and Gd by Balog and Secco.[29] Molodets and Golyshev recently obtained a slight decrease of electrical resistivity of indium at high pressures.[30] We converted the electrical resistance Garg et al. measured[4] into resistivity assuming the length and the width of the sample unchanged in their diamond anvil cell measurements[5] to make direct comparisons with our calculated data. The calculated electrical resistivity of bcc Fe agrees well with experiment at pressures below 10 GPa. Bcc Fe transforms to the hcp structure at 14 GPa at room temperature.[31] Diamond anvil cell experiments show that the electrical resistance of iron has a sharp increase during the phase transition,[4] with the corresponding electrical resistivity much higher than either bcc or hcp phase. Such an abrupt increase in resistivity during the phase transition has also been observed in Ytterbium.[4] At the highest pressure of ~25 GPa in Garg et al's experiment, the measured resistivity is significantly higher than the calculated data, indicating that the initial bcc phase has not fully transformed to hcp structure at this pressure yet, possibly due to pressure gradients. The calculated electrical resistivity of hcp Fe is in close agreement with the LRL demagnetization measurements at ~37.5 GPa,[3] suggesting that most of the bcc phase has already transformed into hcp phase.

**IV.    Conclusions**

In summary, we present the calculated temperature and pressure dependences of the electrical resistivity for bcc and hcp Fe using the low-order variational approximation and theoretical transport spectral functions obtained from first principles linear response linear-muffin-tin-orbital calculations. The calculated values increase nearly linearly with temperature and decrease with atomic volume, in agreement with the Bloch-Grüneisen formula based on first-principles calculated temperature and pressure dependences of the Debye temperature. Overall, first-principles linear response calculations provide an efficient way to examine the temperature and pressure dependences of the transport properties such as electrical resistivity.

**Acknowledgements**

We thank S. Y. Savrasov for kind agreement to use his LMTO codes and many helpful discussions. This work was supported by DOE ASCI/ASAP subcontract B341492 to Caltech DOE w-7405-ENG-48 and by NSF grant EAR-0738061, the Carnegie Institution of Washington, and EFree, an Energy Frontier Research Center funded by the U.S. Department of Energy, Office of Science, Office of Basic Energy Sciences under Award Number DE-SC0001057. Computations were performed at the Geophysical Laboratory and on ALC at Lawrence Livermore National Lab.

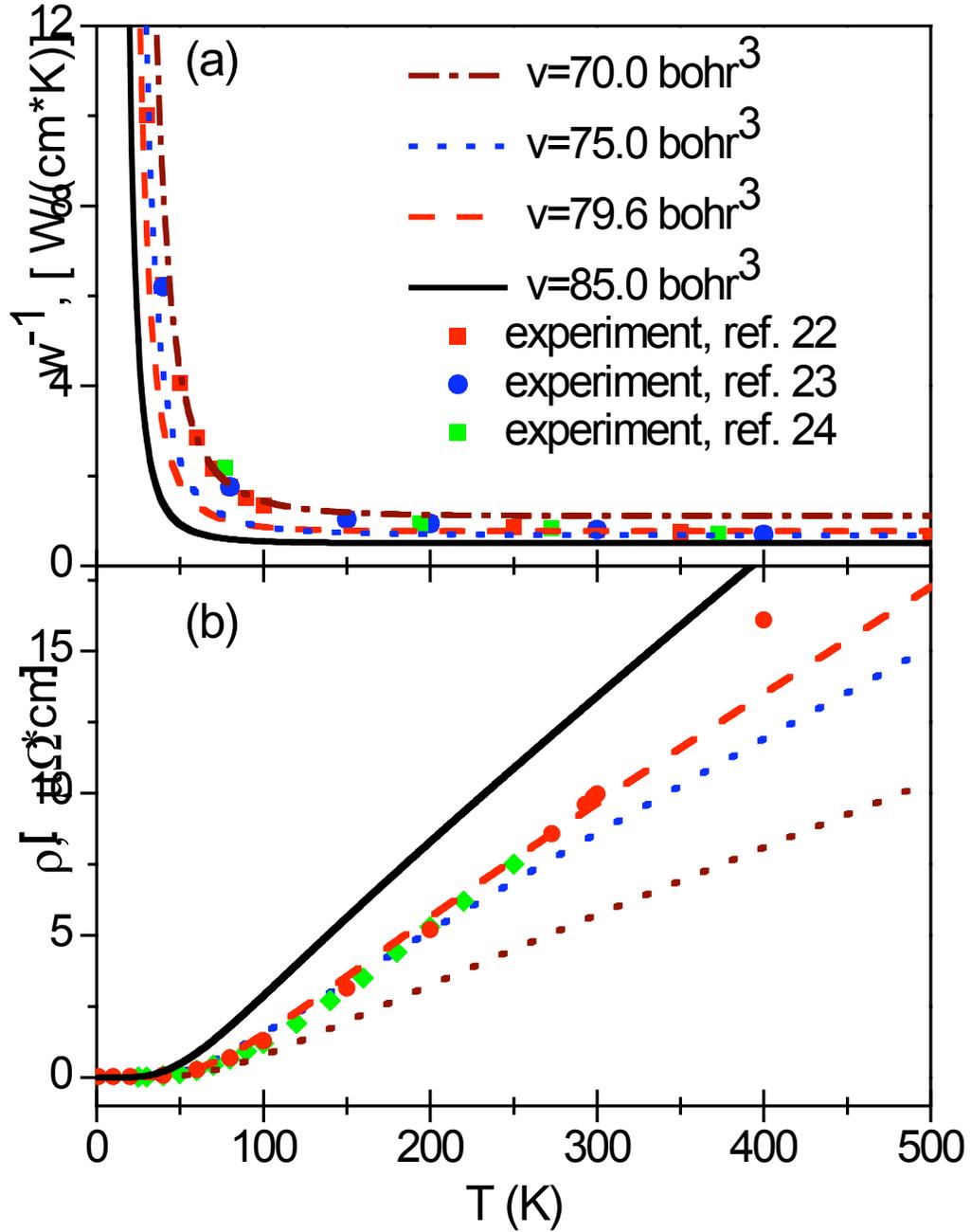

Fig. 1. The calculated temperature dependence of (a) the thermal conductivity, $w^{-1}(T)$ and the electrical resistivity, $\rho(T)$, for bcc Fe at several selected volumes, obtained as a lowest-order variational solution of the Boltzmann equation. Available ambient-pressure experimental data are shown in symbols (Ref. 22-24).

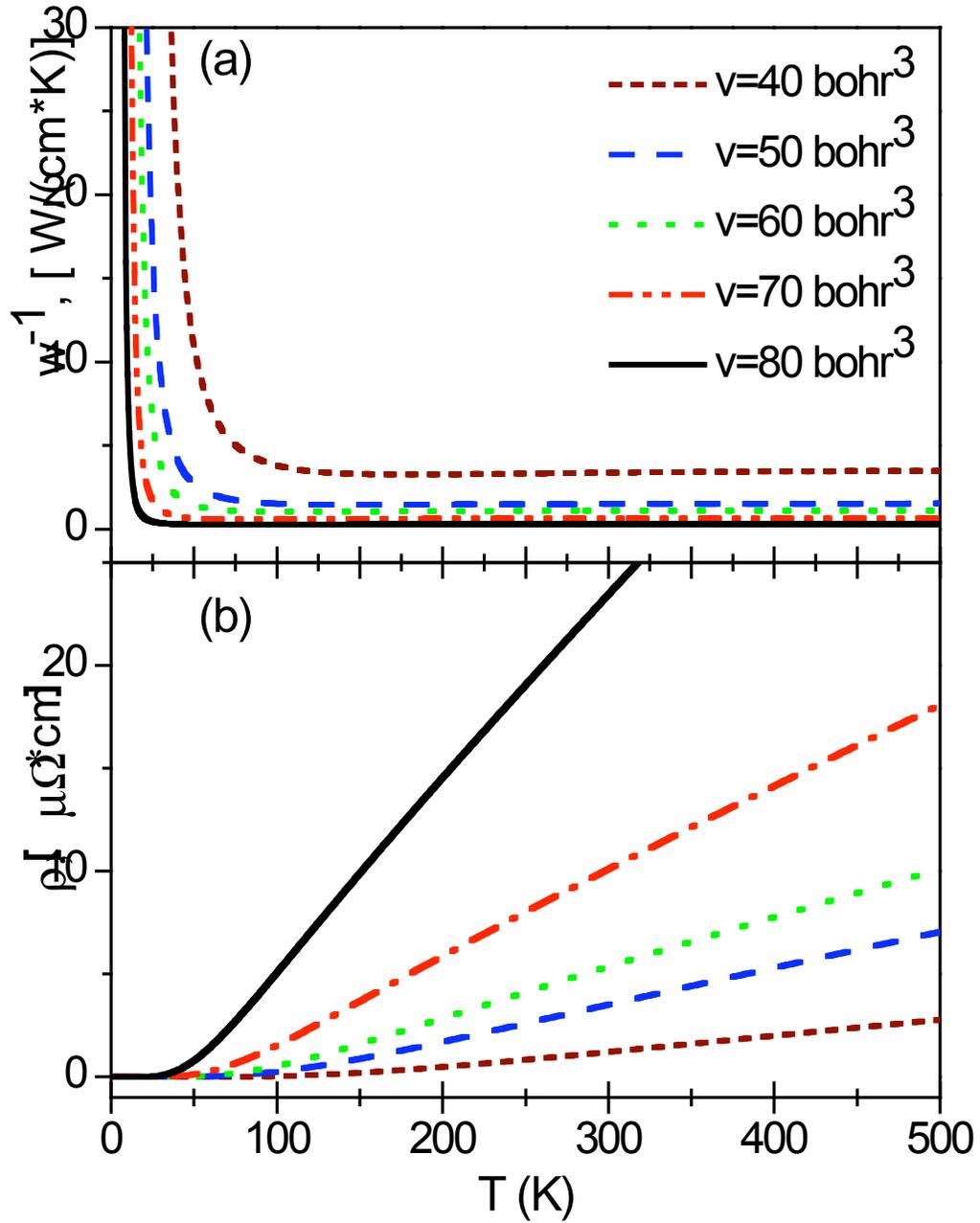

Fig. 2. The calculated temperature dependence of (a) the thermal conductivity, $w^{-1}(T)$ and the electrical resistivity, $\rho(T)$, for hcp Fe at several selected volumes, obtained as a lowest-order variational solution of the Boltzmann equation.

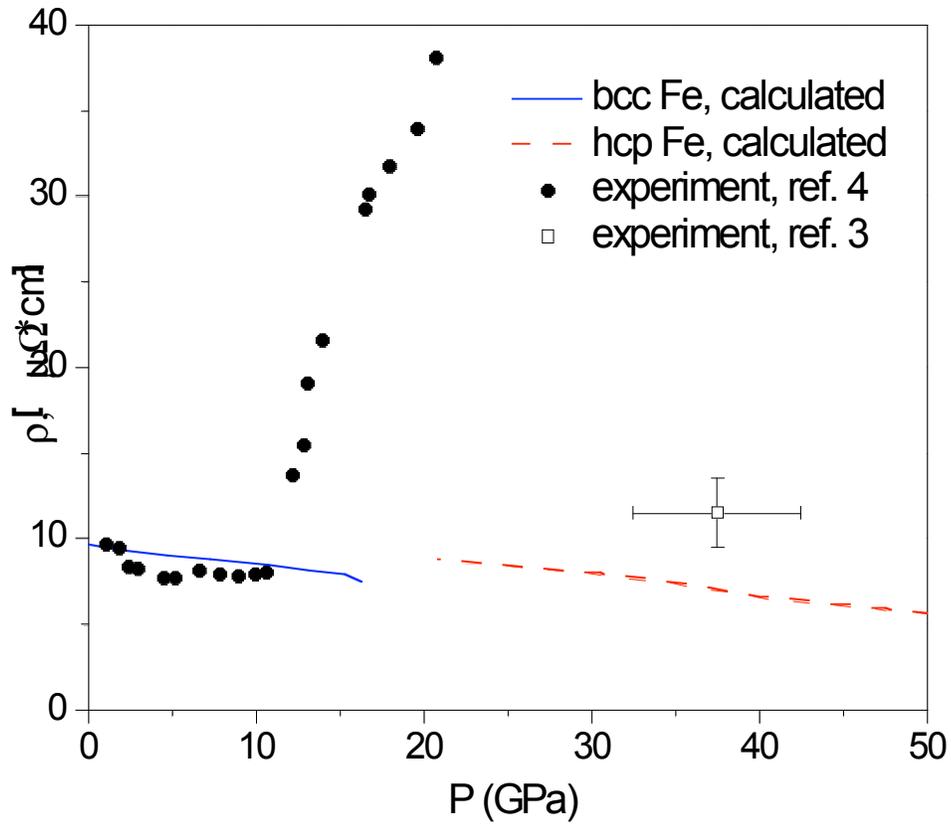

Fig. 3. The calculated pressure dependences of the electrical resistivity for iron at ambient temperature, in comparison to experimental measurements in a diamond anvil cell (Ref. 4) and LRL demagnetization (Ref. 3).

Table I   The calculated values for the transport constant $\lambda_{tr}$ and the average transport frequencies $\Theta_{tr}$ for bcc and hcp Fe at different atomic volumes.

|  | bcc Fe | | | | hcp Fe | | | |
|---|---|---|---|---|---|---|---|---|
| V (bohr$^3$/atom) | 70 | 75 | 79.6 | 85 | 40 | 50 | 60 | 70 |
| $\lambda_{tr}$ | 0.29 | 0.40 | 0.31 | 0.47 | 0.22 | 0.29 | 0.29 | 0.38 |
| $\Theta_{tr}$, K | 389 | 350 | 311 | 260 | 797 | 592 | 468 | 352 |